\input{aipcheck.tex}

\documentclass[
    ,final
  ]
  {aipproc}

\layoutstyle{6x9}

\begin{document}

\title 
      [Gamma rays from galaxy clusters]
      {High energy emission from galaxy clusters and particle acceleration 
due to MHD turbulence}


\author{G. Brunetti}{
  address={INAF Istituto di Radioastronomia, via P.Gobetti 101, 40129,
  Bologna, Italy},
  email={brunetti@ira.inaf.it},
}
\iftrue
\author{P. Blasi}{
address={INAF Osservatorio Astrofisico di Arcetri, Firenze, Italy and
Theoretical Astrophysics, Fermi National Accelerator Laboratory,
Batavia, USA},
}
\iftrue
\author{R. Cassano}{
address={INAF Istituto di Radioastronomia, via P.Gobetti 101, 40129,
    Bologna, Italy},
}
\iftrue
\author{S. Gabici}{
address={Dublin Institute for Advanced Studies, 31 Fitzwilliam Place,
Dublin, Ireland}
}

\fi
\copyrightyear  {2001}

\begin{abstract}
In the next years the FERMI gamma ray telescope and the Cherenkov 
telescopes will put very stringent constraints to models of gamma ray 
emission from galaxy clusters providing crucial information on 
relativistic particles in the inter-galactic-medium.

We derive the broad band non-thermal spectrum of galaxy clusters 
in the context of general calculations in which relativistic particles 
(protons and secondary electrons due to proton-proton collisions) interact 
with MHD turbulence generated in the cluster volume during cluster mergers,
and discuss the importance of future gamma ray observations.
\end{abstract}

\date{\today}

\maketitle

\section{Introduction}

Clusters of galaxies contain $\approx 10^{15}$ M$_{\odot}$ of 
hot ($10^8$ K) gas, galaxies, dark matter and non-thermal components.  
Theoretically, non-thermal components such as magnetic fields and high 
energy particles play key roles by controlling transport processes in the 
inter-galactic-medium (IGM) and are sources of additional pressure and 
energy support \cite{ryu03, narayan, lazarian06, gb+laz07, ando}. 
Therefore they have the potential to invalidate our simplified view of 
the properties of the IGM.

The origin of non-thermal components is likely connected with the cluster 
formation process : it is believed that 
a fraction of the energy dissipated during cluster 
mergers and accretion of matter is channelled into the 
acceleration of particles via shocks and turbulence that lead to a complex 
population of primary electrons and protons in the IGM (e.g., \cite{enss98, 
sara99, blasi01, gb01, gabici03, gb04, petro01, min01, ryu03, gb+laz07, pf08}).
Relativistic protons are expected to be the dominant non-thermal particle 
components since they have long life-times and remain confined within galaxy 
clusters for an Hubble time (e.g. \cite{blasi07} and ref. therein). 
Collisions between relativistic protons and thermal protons in the IGM
inject secondary electrons and neutral pions that in turns produce synchrotron 
(and inverse Compton, IC) and gamma-ray emission, respectively, 
whose relevance depends on the relativistic-proton content in the IGM 
\cite{voelk96, bere97, blasi99, min01, pfrommer+ensslin04}.

The presence of non-thermal particles in the IGM is 
proved by radio observations that discovered Mpc-sized diffuse radio emission 
in a fraction of X-ray luminous galaxy clusters due to synchrotron radiation 
from relativistic electrons in the magnetised IGM.
These diffuse radio sources are usually classified according to their 
morphology and connection with the cluster X-ray emitting gas: Radio Halos, 
fairly symmetric radio structures at cluster center, and Radio Relics, 
elongated sources at cluster periphery (e.g. \cite{feretti03, ferrari08}).

Giant Radio Halos are the most prominent examples of the diffuse cluster 
non-thermal emission. 
These Mpc-scale sources may originate from {\it secondary electrons}
injected by collisions between relativistic and thermal protons in
the IGM (e.g. \cite{dennison80, blasi99}), or they may originate from 
relativistic electrons {\it re-accelerated} {\it in situ} by various 
mechanisms associated with the turbulence in massive merger events 
(e.g. \cite{gb01, petro01, fujita03}).
The recent discovery of a Radio Halo with very-steep synchrotron spectrum in 
Abell 521 supports the turbulent acceleration mechanism \cite{gb08} providing 
a glimpse of what the next generation of low-frequency radio telescopes
might find, since the turbulent acceleration scenario predicts that many Radio 
Halos in the Universe should emit mainly at low radio frequencies 
\cite{cassano+gb+setti06}.

Theoretically particle acceleration by MHD turbulence may
happen at the same time with the unavoidable continuous injection of
secondary electrons due to proton-proton collisions in the IGM, and the 
non-thermal properties of the IGM that come out from the interplay between 
these two processes mainly depends on the content of relativistic protons 
in the IGM \cite{gb+blasi05}.

The properties of relativistic protons in the IGM are still poorly constrained 
since present gamma ray observations can provide only upper limits to the 
gamma ray emission from galaxy clusters (e.g., \cite{reimer03}; see Sect.2).
However, the FERMI gamma-ray space telescope will shortly allow to 
measure (or constrain) the energy content of relativistic protons in the IGM.
For this reason in this contribution 
we will use a general approach, in which relativistic particles
(protons and secondary electrons due to proton-proton collisions) interact 
with MHD turbulence, and show expectations for gamma ray emission from 
galaxy clusters and its connection with the emission at other wavelengths.
We will not consider the contribution to gamma ray emission due to other 
possible relevant processes, such as IC emission from high energy electrons
accelerated at peripheral strong shock waves and dark matter annihilation 
\cite{loeb00, miniati03, gabici04, blasi07, pf08, profumo}.

In Section 2 we briefly discuss available constraints on the
content of relativistic protons in galaxy clusters and in Section 3 we
discuss the non-thermal spectrum of galaxy clusters and the importance
of future gamma ray observations with FERMI and Cherenkov telescopes.

\noindent
We adopt concordance $\Lambda$CDM cosmology with $H_o=70$ km
s$^{-1}$ Mpc$^{-1}$, $\Omega_m=0.3$ and $\Omega_{\Lambda}=0.7$.

\section{Present limits on cosmic ray protons in the IGM}

A fairly natural consequence of our present theoretical view of cosmic rays 
in galaxy clusters is that relativistic protons should be the dominant 
non-thermal particles component in the IGM, and that their properties trace 
the history of the complex interplay between particle acceleration 
and advection processes that take place in galaxy clusters from their 
formation epoch (e.g. \cite{blasi07} and ref. therein). 
Remarkably, if the energy density of these non-thermal protons is more than 
a few percent of that of the thermal IGM, protons are also relevant to
understand the physics and the origin of the relativistic electrons 
responsible for the synchrotron emission in the form of
Radio Halos \cite{gb+blasi05}.

The most direct approach to constrain the energy content of
relativistic protons in galaxy clusters consists in the observation of 
the clusters' gamma ray emission from the decay of the neutral pions 
that originate
during proton-proton collisions in the IGM.

\noindent
Gamma ray upper limits from EGRET observations 
allow to constrain the energy density of non-thermal protons in nearby
galaxy clusters at less than about 30 percent of the thermal IGM 
\cite{reimer03, pfrommer+ensslin04}. Remarkably, since these limits are 
obtained from observations at energies of about 100 MeV, they are
fairly independent of the spectral shape of the energy distribution of
relativistic protons.

\noindent
More recently, deep gamma ray observations with Cherenkov telescopes of 
a few nearby galaxy clusters allow to put more stringent constraints to the
content of relativistic protons in the IGM. These limits are obtained
from observations at $>$100 GeV and thus they depend on the spectral shape of
the proton-energy distribution. Yet, in the relevant case in which 
the spectrum of relativistic protons is relatively flat and
their spatial distribution follows that of the IGM, these Cherenkov 
observations constrain the energy density of relativistic protons 
at $<$10\% of the thermal IGM (\cite{perkins06, aharonian08}; Figure 1).

Also radio observations of galaxy clusters can be used to constrain the 
energy content of relativistic protons in the IGM
\cite{reimer04, gb07, gb08}.
In these studies radio upper limits to the diffuse cluster-scale
synchrotron emission due to secondary electrons are used to constrain 
the energy density of the primary relativistic protons.
Limits based on this approach depend also on the cluster magnetic field 
strength in the IGM and
are complementary to the limits obtained from gamma ray observations.
However, due to the high sensitivity of present radio telescopes, radio 
observations of a fairly large number of galaxy clusters without Radio Halos
allow to put very stringent limits to the energy density of non-thermal 
protons, $E_{CR}/E_{IGM} \leq 0.01$, provided that the strength of the
cluster-scale magnetic field is $> \mu$G (Figure 1; \cite{gb07, venturi08}).

\begin{figure}
\caption{
Present limits to the ratio between relativistic protons 
and IGM energy densities.
The thick--dashed upper limit indicates approximatively the limit to 
the ratio $E_{CR}/E_{IGM}$ obtained from Cherenkov observations 
(e.g., \cite{perkins06} for the Perseus clusters and 
\cite{aharonian08} for Abell 85). Solid upper limits give the typical
limits to the ratio $E_{CR}/E_{IGM}$ obtained from GMRT radio observations 
of a sample of X-ray selected clusters with no evidence of Radio Halos 
\cite{gb07, venturi08}; limits based on radio observations depend on the
magnetic field strength, $B$, in the central Mpc region.
Limits are obtained assuming that the spatial distribution of non-thermal
protons follows that of the thermal IGM and assuming a relatively flat
spectral energy distribution of relativistic protons, $N(p)\propto
p^{-\delta}$, with $\delta = 2.2$. }
\includegraphics[height=.45\textheight]{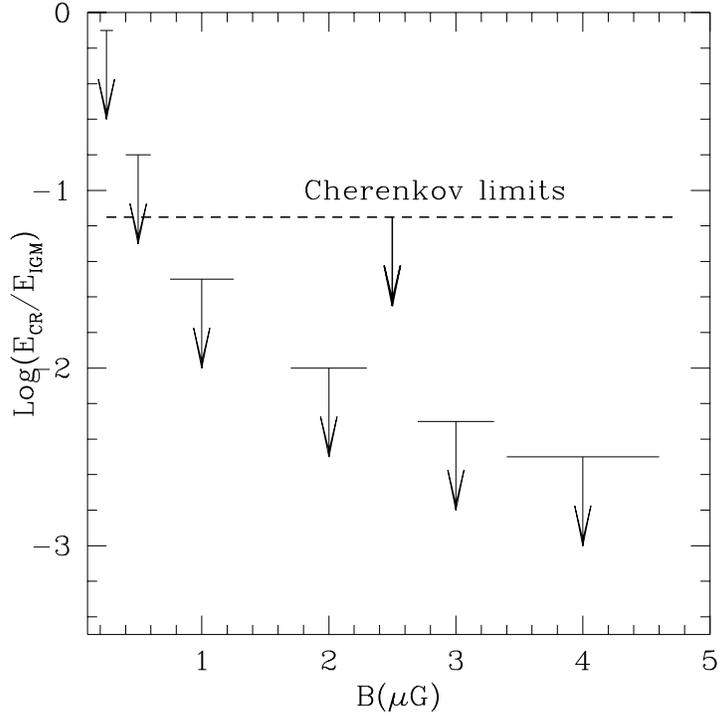}
\end{figure}

\section{Non-thermal emission from galaxy clusters}

In this Section we derive the non-thermal spectrum of galaxy clusters under 
the assumption that MHD turbulence is injected in the IGM in connection 
with cluster mergers and plays a role in the particle acceleration process 
during these mergers.

\noindent
We restrict to the case in which the energy density of relativistic protons 
in galaxy clusters 
is consistent with the constraints discussed in the previous Section and 
model the re-acceleration of relativistic particles by MHD turbulence 
in the most simple situation in which only these relativistic protons are 
initially present in the IGM. 
Protons generate secondary electrons via proton-proton collisions and 
in turns secondaries (as well as protons) are re-accelerated by 
MHD turbulence.

\noindent
The non-thermal spectrum emitted in this general scenario is complex
\cite{gb+blasi05, gbgb}.
Secondary particles and the decay of neutral pions continuously 
generated by proton-proton collisions in the IGM generate a long-living
emission, from radio to gamma-rays.
The strength and spectral properties of this emission depend on the 
energy density (and spectrum) of non-thermal 
protons in the IGM (and on the magnetic field in the case of the 
synchrotron radio emission), and is almost independent of the dynamics of
the cluster.

\noindent
In addition, particle re-acceleration due to MHD turbulence that may 
happen in connection with cluster mergers can
modify the shape of the spectral energy distribution 
of the relativistic particles and consequently 
that of the cluster non-thermal emission.

\noindent
The most important issues that we would like to address in this
Section are :

\begin{itemize}

\item
the possibility to have a {\it radio loud phase}, during which Radio
Halos are generated in galaxy clusters, and a {\it radio quiet phase}, 
during which the diffuse synchrotron emission from clusters is significantly 
smaller than that of presently known Radio Halos;

\item
the level of the cluster gamma ray emission expected from those models that 
may successfully reproduce a {\it radio loud} and a {\it radio quiet phase}
in clusters (and possibly also a transition between the two phases).

\end{itemize}

It is important to mention that the first point is motivated by 
radio observations of X-ray selected 
cluster samples that demonstrate that Radio Halos are not common in galaxy
clusters \cite{gg99,buote01, cassano08, venturi08} and that the synchrotron 
luminosity of clusters with no evidence of Radio Halos 
({\it radio quiet phase}) is (at least) about one order of magnitude 
fainter than that of Radio Halos ({\it radio loud phase}) associated with 
clusters with similar X--ray luminosity and redshift \cite{gb07, venturi08}.
In addition, the rarity of clusters with diffuse cluster-scale emission
at intermediate level, i.e. between that of Radio Halos 
and of {\it radio quiet} clusters,
suggests that the time-scale necessary to switch-on (or to switch-off) 
these Radio Halos is fairly short, $\approx 0.1-0.3$ Gyrs \cite{gb07}
implying that the bulk of relativistic electrons in the IGM is only 
maintained for such a limited period in connection with cluster 
mergers \cite{gb07}
\footnote{an additional possibility is that the transition between
{\it radio quiet} and  {\it radio loud phases} is due to fast dissipation
of the cluster magnetic field after cluster mergers. 
However, the problem here is that
even in the {\it most favorable case} in which the magnetic 
field is simply dissipated during the decay of cluster-MHD turbulence, the 
energy density of the rms field decreases only (about) linearly
with the eddy turnover time-scale requiring $\approx$ Gyr for 
(substantial) dissipation (e.g., Fig.2 in \cite{subramanian06})}.

Following \cite{gb+blasi05} we calculate particle acceleration due to
MHD turbulence during mergers by restricting to the case of
Alfven waves\footnote{an additional possibility is given
by magnetosonic waves \cite{cassano+gb05, gb+laz07}}
and calculate the spectrum of particles 
and MHD waves and their evolution with time by solving a set
of coupled equations
that give the spectrum of electrons, $N_e^-$, positrons, $N_e^+$,
protons, $N_p$, and waves, $W_k$ :

\begin{eqnarray}
{{\partial N_e^{\pm}(p,t)}\over{\partial t}}=
{{\partial }\over{\partial p}}
\Big[
N_e^{\pm}(p,t)\Big(
\left|{{dp}\over{dt}}_{\rm r}\right| -
{1\over{p^2}}{{\partial }\over{\partial p}}(p^2 D_{\rm pp}^{\pm})
+ \left|{{dp}\over{dt}}_{\rm i}
\right| \Big)\Big] + 
\nonumber\\
{{\partial^2 }\over{\partial p^2}}\left[
D_{\rm pp}^{\pm} N_e^{\pm}(p,t) \right] + Q_e^{\pm}[p,t;N_p(p,t)] \, ,
\label{elettroni}
\end{eqnarray}

\begin{eqnarray}
{{\partial N_p(p,t)}\over{\partial t}}=
{{\partial }\over{\partial p}}
\Big[
N_p(p,t)\Big( \left|{{dp}\over{dt}}_{\rm i}\right|
-{1\over{p^2}}{{\partial }\over{\partial p}}(p^2 D_{\rm pp})
\Big)\Big]
+ {{\partial^2 }\over{\partial p^2}}
\left[ D_{\rm pp} N_p(p,t) \right] \, ,
\label{protoni}
\end{eqnarray}

and 

\begin{eqnarray}
{{\partial W_{\rm k}(t)}\over
{\partial t}} =
{{\partial}\over{\partial k}}
\left( k^2 D_{\rm kk} {{\partial}\over{\partial k}} \left[
{{W_k(t)}\over{k^2}} \right] \right)
- \Gamma(k) W_{\rm k}(t) 
+ I_{\rm k}(t), 
\label{turbulence}
\end{eqnarray}

where $|dp/dt|$ marks radiative (r) and Coulomb (i) losses, 
$D_{pp}$ is the particle diffusion coefficient in the momentum
space (and depends on the wave spectrum 
$W_{\rm k}$), $Q_{e}^{\pm}$ is
the injection term of secondary leptons due to proton-proton collisions 
(and depends on $N_p$), $D_{kk}$ is the diffusion coefficient in 
the wavenumber space,
$I_k$ is the injection rate-spectrum of Alfven waves at resonant
scales, and $\Gamma$ is the damping rate of waves due to non-linear
resonance with thermal and relativistic particles ($N_p$ and $N_e$);  
details can be found in \cite{gb+blasi05}.

\begin{figure}
  \resizebox{18pc}{!}{\includegraphics{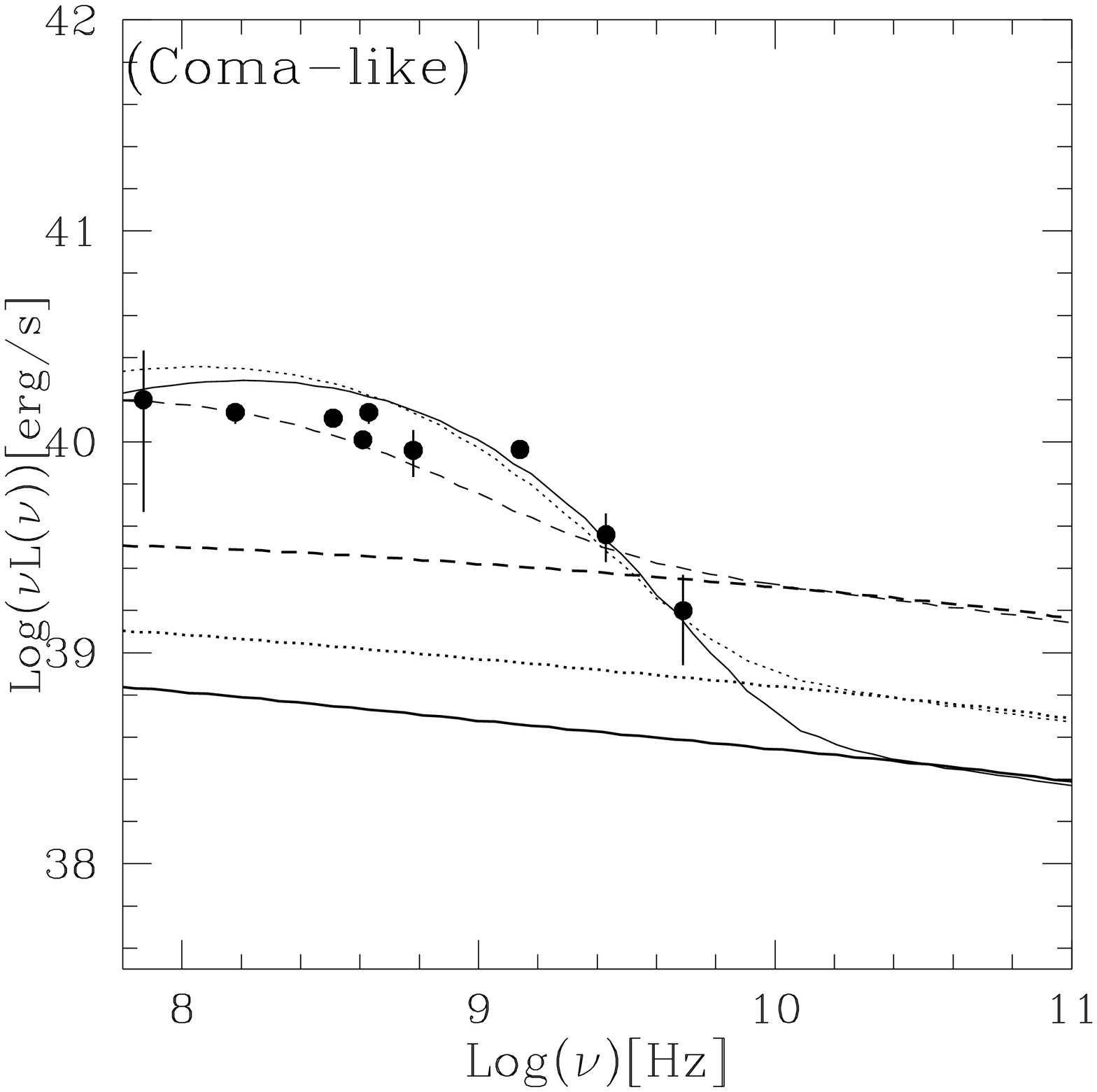}}
  \resizebox{18pc}{!}{\includegraphics{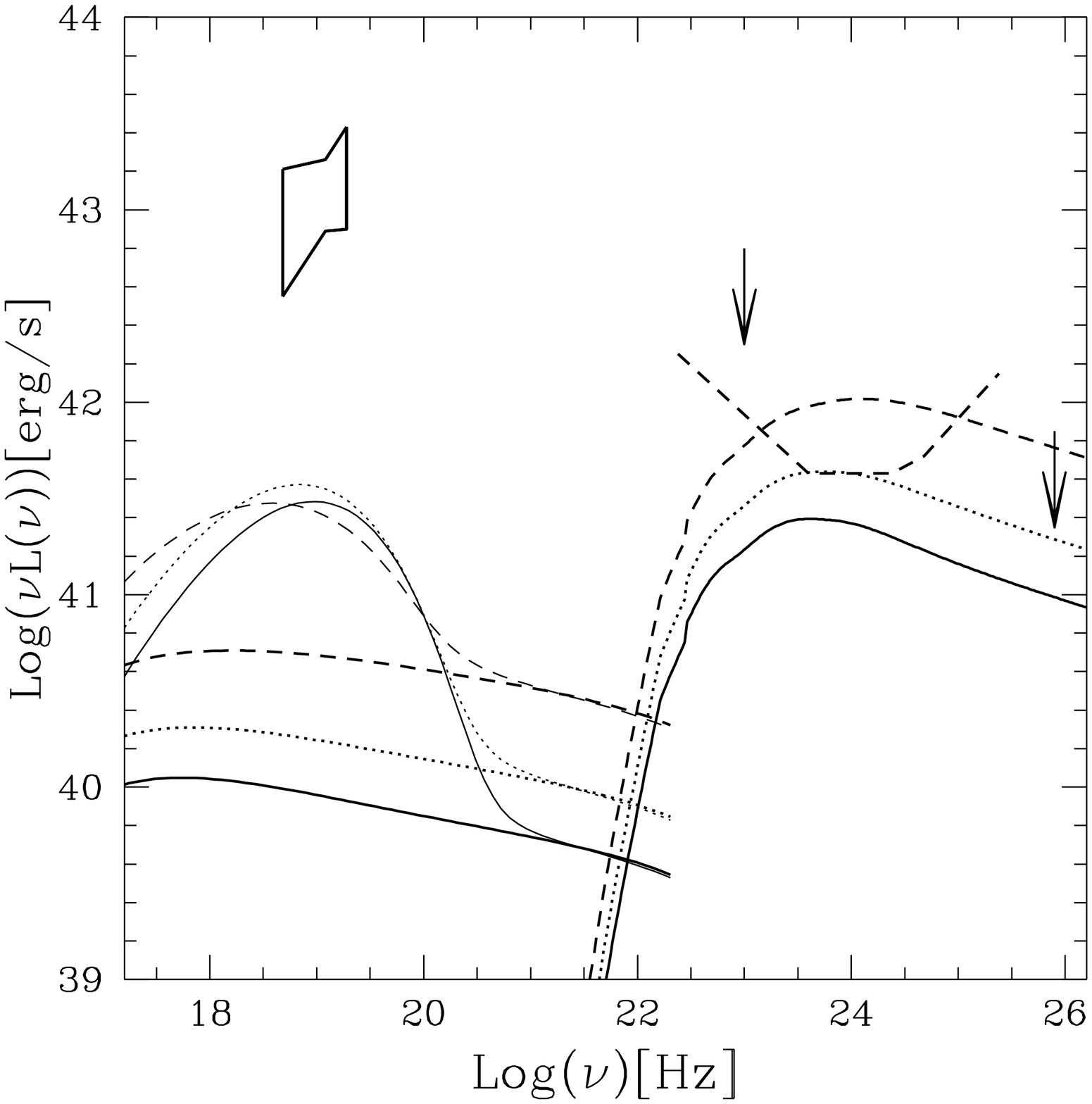}}
\caption{
Broad band spectrum produced within 3.5 core radii from a Coma-like cluster.
\noindent
Synchrotron (SZ decrement at high frequencies is not taken into account)
(left panel) and IC and pion-decay spectra (right panel)
calculated at t=0.5 Gyr from the injection of small scale Alfven waves 
in the IGM (the energy injected in these waves between t=0-0.5 Gyr 
is $\approx$3\% of the thermal IGM) (thin lines) and calculated at t=1 Gyr 
after dissipation of turbulence (thick lines).
\noindent
Calculations are shown assuming a ratio between the energy density of
relativistic and thermal protons at the beginning of the merger phase
= 3\% (dashed lines), 1\% (dotted lines) 
and 0.5\% (solid lines) at t=0 (with proton spectrum $\delta
=2.2$) and a central cluster-magnetic field $B_o=2 \mu$G.
\noindent
Radio data, BeppoSAX data and EGRET upper limit for the Coma cluster 
(\cite{gb+blasi05} and ref. therein) are shown together with the recent 
VERITAS upper limit (calculated starting from \cite{perkins08}) and
with the FERMI approximate sensitivity, after 1 yr, at distance of
the Coma cluster (dashed curve). 
}
\end{figure}

We adopt a simple model of galaxy clusters by assuming that the 
energy densities of relativistic protons at the beginning of 
re-acceleration, $E_{CR}$, and of the magnetic field, $B$, and the 
injection rate of Alfven waves during mergers, $I_{k}$, scale with thermal 
energy density, $E_{IGM}$ ($E_{CR} \propto E_{IGM}$, $B \propto E_{IGM}$ 
and $\int I_{k} dk \propto E_{IGM}$).
Finally we model the spatial distribution of the thermal IGM with a 
spherically symmetric beta-model with the parameters of the 
Coma cluster. We solve Eqs.1--3 at different distances from cluster center
and calculate the total non-thermal broad-band radiation (synchrotron, IC, 
$\pi^o$ decay) emitted from the cluster volume (Figure 2).

To highlight the effect of particle re-acceleration by MHD turbulence
during cluster mergers, in Figure 2 we show non-thermal spectra emitted
in connection with a merger and 1 Gyr after turbulence in the IGM
is dissipated. 
Remarkably, the results have the potential to explain cluster
{\it radio loud} and {\it radio quiet phases} : Radio Halos develop in 
connection with particle re-acceleration due to MHD turbulence in cluster 
mergers where the cluster-synchrotron emission is considerably
boosted up, while a fainter long-living radio emission from 
secondary electrons is expected to be common in clusters.
The strength of this latter component is proportional to the energy density
of relativistic protons in the IGM and is constrained by the 
upper limits to the cluster-scale radio emission in galaxy clusters 
without Radio Halos that are available from present 
radio observations \cite{gb07, venturi08}.

\noindent
IC hard X-ray emission is also boosted up in connection with Radio Halos, 
although the IC signal from re-accelerated {\it secondary} electrons 
is not expected to be very luminous 
(see discussion in \cite{gb+blasi05} for a comparison with the
case of re-acceleration of {\it primary relic} electrons).

An important point is that gamma ray emission is expected (at some level, 
depending on the content and spatial distribution of relativistic protons) 
to be common in galaxy clusters and not directly correlated with the 
presence of giant Radio Halos.
Cherenkov arrays already constrain the level of these gamma rays from a few
nearby clusters. Most importantly, in Figure 2 we show the preliminary
upper limit obtained from a deep VERITAS observation of the Coma 
cluster \cite{perkins08} that, if confirmed, is 
the most stringent constraint to the energy density of relativistic 
protons in a cluster of galaxies obtained so far by means of 
gamma ray observations; based on the results reported in Figure 2 this
limit is about $E_{CR}/E_{IGM} < 0.05$.

\noindent
In Figure 2 we also show the approximate level of the sensitivity of 
the FERMI gamma ray telescope, after $\approx$1 yr of operations, for
a cluster at the distance of Coma.
Based on the VERITAS limit, FERMI has still a chance to marginally
detect gamma ray emission produced by the decay of neutral pions in
the Coma cluster after the first year of operations, 
even in the case of a relatively flat spectral energy distribution of 
relativistic protons.

\noindent
Most importantly, in the next years FERMI will reach the 
sensitivity that is necessary to constrain those models that have 
the potential to explain the behavior of the occurrence of Radio Halos 
in galaxy clusters as constrained by present radio observations.

\section{Summary}

Relativistic protons are expected to be the dominant particle
non-thermal component in galaxy clusters. 
On the other hand, the energy density of this component has been 
poorly constrained by past EGRET observations.

\noindent
In these years, 
important constraints to the energy density of 
relativistic protons in the IGM are coming from Cherenkov observations of
a few nearby galaxy clusters. After about 1 year of operations 
the FERMI gamma ray telescope will be able to put even more stringent 
constraints, especially in the case of steeper spectra
of the relativistic protons.

Recent radio observations support the hypothesis, put forward in the
last years, that MHD turbulence
injected in galaxy clusters during cluster-cluster mergers may play
a role in the re-acceleration of the relativistic electrons responsible
for the cluster-scale synchrotron emission detected in a fraction
of merging and massive galaxy clusters.

\noindent
We derive the non-thermal spectrum from galaxy clusters by
considering the non-linear coupling between MHD turbulence, driven in 
the IGM during cluster mergers, relativistic protons and secondary
electrons.
Radio (synchrotron) and X-ray (IC) emission, due to secondary 
electrons, and gamma ray emission, due to the decay of neutral pions, 
are expected to be common in galaxy clusters.
In addition to this long-living component we have shown that particle
re-acceleration due to MHD turbulence, in connection with cluster mergers, 
can considerably boost up the synchrotron and IC components.

\noindent
The overall picture has the potential to explain the well established 
connection between
Radio Halos and cluster mergers, and the lack of detectable
cluster-scale radio emission in the majority of galaxy clusters 
with X--ray luminosity and redshift similar to those of Radio-Halo clusters.

\noindent
We point out that in the next years the FERMI telescope will efficiently 
constrain these models and may have the chance to detect 
gamma rays from nearby massive clusters (e.g Coma), provided that the 
energy density of relativistic protons is (at least) 1--3 percent of the 
thermal IGM.

\noindent
Remarkably, for relatively flat spectra of the relativistic protons, 
as expected in the 
case these protons got a sizeable fraction of the energy of the thermal 
IGM trough efficient particle acceleration processes, extremely deep 
($\geq 100$ hrs) future observations with Cherenkov telescopes 
may obtain the most stringent constraints.

\begin{theacknowledgments}
We acknowledge partial support from ASI-INAF I/088/06/0 and 
PRIN-INAF2007.
\end{theacknowledgments}

\bibliographystyle{aipprocl} 

{}

\end{document}